\documentclass[]{sig-alternate-05-2015}
\usepackage{bm}
\usepackage{bbm} 
\usepackage{framed}
\usepackage{algorithm}
\usepackage{algorithmic}
\usepackage{overpic}
\newtheorem{lemma}{Lemma}
\newtheorem{theorem}{Theorem}
\newtheorem{definition}{Definition}
\newtheorem{proposition}{Proposition}
\begin{document}

\title{Throughput-Optimal Multi-hop Broadcast Algorithms}

\numberofauthors{3}
    \author{
      \alignauthor Abhishek Sinha\\ 
      \affaddr{Laboratory for Information and Decision Systems}\\
      \affaddr{MIT}     
      \email{sinhaa@mit.edu}
  \and \and     \alignauthor Georgios Paschos\\
      \affaddr{Mathematical and Algorithmic Sciences Lab }
      \affaddr{Huawei Technologies Co. Ltd} \\    
   \email{ georgios.paschos@huawei.com}
 \and \and 
     \alignauthor Eytan Modiano\\  
     \affaddr{Laboratory for Information and Decision Systems}\\
      \affaddr{MIT}     
      \email{modiano@mit.edu}  
%
          }
\maketitle

\begin{abstract}
	In this paper we design throughput-optimal dynamic broadcast algorithms for multi-hop networks with arbitrary topologies. Most of the previous broadcast algorithms route packets along spanning trees, rooted at the source node. For large dynamic networks, computing and maintaining a set of spanning trees is not efficient, as the network-topology may change frequently. In this paper we design a class of dynamic algorithms which makes packet-by-packet scheduling and routing decisions and thus obviates the need for maintaining any global topological structures, such as spanning trees. Our algorithms may be conveniently understood as a non-trivial generalization of the familiar \emph{back-pressure} algorithm which makes unicast packet routing and scheduling decisions, based on queue-length information, without maintaining end-to-end paths. However, in the broadcast problem, it is hard to define queuing structures due to absence of a work-conservation principle which results from packet duplications. We design and prove the optimality of a virtual-queue based algorithm, where a virtual-queue is defined for subsets of vertices. We then  propose a multi-class broadcast policy which combines the above scheduling algorithm with a class-based \emph{in-order} packet delivery constraint, resulting in significant reduction in complexity. Finally, we evaluate performance of the proposed algorithms via extensive numerical simulations.   
\end{abstract}
\section{Introduction} \label{introduction_section}
Packet broadcasting is used for efficiently disseminating messages to all recipients in a network. Its efficiency is measured in terms of \emph{broadcast throughput}, i.e., the common rate of packet-reception by all  nodes. Technically, the broadcast problem refers to finding a policy for duplicating and forwarding copies of packets such that the maximum broadcast throughput (also known as \emph{broadcast-capacity}) is achieved.\\
Solving the broadcast problem is challenging, especially for mobile wireless networks with time-varying connectivity and interference constraints. In this paper we focus on designing dynamic broadcast algorithms. Such algorithms operate without the knowledge of network-topology or future arrivals, and hence, are robust. In this context, we derive provably throughput-optimal dynamic broadcast algorithms for  networks with arbitrary topology.

Most of the existing broadcast algorithms are static by nature and operate by forwarding copies of packets along spanning trees \cite{swati}. In a network with time-varying topology, these static algorithms need to re-compute the trees every time the underlying topology changes, which could be quite cumbersome and inefficient. Recent works \cite{sinha_DAG} and \cite{rate} consider the problem of throughput-optimal broadcasting in Directed Acyclic Graphs (DAG). Here the authors propose dynamic policies by exploiting the properties of DAG. However, it is not clear how to extend their algorithms to networks with arbitrary (non-DAG) topology. The authors in \cite{massoulie2007randomized} propose a randomized packet-forwarding policy for wireline networks, which is shown to be throughput-optimal under some assumptions. However, their algorithm  potentially needs to use unbounded amount of memory and can not be easily generalized to wireless networks with activation constraints. A straight-forward extension of their algorithm, proposed in \cite{towsley2008rate}, uses activation oracle, which is not practically feasible.\\   
In this paper we study the broadcasting problem in arbitrary networks, including wireless. We propose algorithms that do not require the construction of global topological structures, like spanning trees.  Leveraging the work in \cite{sinha_DAG}, we propose a novel multi-class heuristic, which simplifies the operational complexity of the proposed algorithm. Our main technical contributions in this paper are as follows:
	 (1) We first identify a state-space representation of the network-dynamics, in which the broadcast-problem reduces to a ``virtual-queue" stability problem. By utilizing techniques from Lyapunov-drift methodology, we derive a throughput-optimal broadcast policy.
	 (2) Next, we introduce a multi-class heuristic policy, by combining the above scheduling rule with \emph{in-class in-order} packet delivery, where the number of classes is a tunable parameter, which may be used as a trade-off between efficiency and complexity. (3) Finally, we validate the theoretical ideas through extensive numerical simulations. (4) An equivalent \emph{mini-slot} model is proposed, which simplifies the analysis and may be of independent theoretical interest.\\
The rest of the paper is organized as follows. In section \ref{system_model} we describe the operational network model and characterize its broadcast-capacity. In section \ref{optimal_policy} we derive our throughput-optimal broadcast policy. In section \ref{heuristic_section} we propose a multi-class heuristic policy which uses the scheduling scheme from section \ref{optimal_policy}. In section \ref{simulation_section} we validate our theoretical results via extensive numerical simulations. Finally in section \ref{conclusion_section} we conclude the paper with some directions for future work.

\section{System Model} \label{system_model}
For simplicity, we first consider the problem in a wireline setting. The wireless model will be considered in section \ref{int_cont}.
\subsection{Network Model} Consider a graph $\mathcal{G}(V,E)$, $V$ being the set of vertices and $E$ being the set of edges, with $|V|=n$ and $|E|=m$. Time is slotted and the edges are directed. Transmission capacity of each edge is one packet per slot. External packets arrive at the source node $\texttt{r} \in V$. The arrivals are i.i.d. at every slot with expected arrival of $\lambda$ packets per slot. \\
For sake of convenience, we alter the slotted-time assumption and adopt a slightly different but equivalent \emph{mini-slot} model. A slot consists of $m$ consecutive mini-slots. As will be evident from what follows, our dynamic broadcast algorithms are conceptually easier to derive, analyze and understand in the mini-slot model. However, the algorithms can be easily applied in the more traditional slotted model.\\
\emph{Mini-slot model:} In this model, the basic unit of time is called a \emph{mini-slot}. At each mini-slot $t$, an edge $e=(a,b)\in E$ is chosen for activation, independently and uniformly at random from the set of all $m$ edges. All other $m-1$ edges remain idle for that mini-slot. A packet can be transmitted over an active edge only. A single packet transmission takes one mini-slot for completion. This random edge-activity process is represented by the i.i.d. sequence of random variables $\{S(t)\}_{t=1}^{\infty}$, such that, if an edge $e \in E$ is chosen for activation at the mini-slot $t$, we have $S(t)=e$. Thus,
\begin{eqnarray*}
	\mathbb{P}(S(t)=e)=1/m, \hspace{10pt} \forall e \in E, \hspace{5pt} \forall t  
\end{eqnarray*}
External packets arrive at the source $\texttt{r}$ with expected arrival of $\lambda/m$ packets per mini-slot.

The main operational advantage of the mini-slot model is that only a single packet transmission takes place at a mini-slot, which makes it easier to express the system-dynamics. However, as we show in Lemma \eqref{invariance}, these two models are equivalent from the point-of-view of broadcast-capacity.

\subsection{Broadcast-Capacity of a Network}
 Informally, a network supports a broadcast-rate of $\lambda$ if external packets arrive at the source at the rate of $\lambda$ and there exists a scheduling policy under which all nodes receive distinct packets at the rate of $\lambda$. The broadcast-capacity $\lambda^*$ is the maximally achievable broadcast-rate in the network.

Formally, we consider a class $\Pi$ of scheduling policies which executes the following two actions at every mini-slot $t$ 
\begin{itemize}
	\item The policy observes the currently active edge $e=(a,b)$. 
	\item The policy transmits (at most) one packet from node $a$ to node $b$ over the active edge $e$.
	\end{itemize}
	 The policy-class $\Pi$ includes policies that have access to all past and future information, and may forward any packet present at node $a$ at time $t$ to  node $b$.\\
  Recall that, a slot consists of $m$ consecutive mini-slots. Let $R_i^{\pi}(t)$ be the number of distinct packets received by node $i \in V$ up to slot $t$, under a policy $\pi\in \Pi$. The time average $\liminf_{T\to \infty} R^{\pi}_i(T)/T$ is  the rate at which distinct packets are received at  node $i$, under the action of the policy $\pi$.
 
\begin{definition}[Broadcast Policy]
A policy $\pi$ is called a 
{``broadcast policy of rate $\lambda$''} 
if 
all nodes in the network receive distinct packets at the rate of $\lambda$ packets per slot, i.e.,
\begin{eqnarray} \label{bcdef}
\min_{i\in V} \liminf\limits_{T\to \infty} \frac{1}{T} R^{\pi}_i(T)= \lambda, \hspace{10pt} \mathrm{in \hspace{2pt} probability }, 
\end{eqnarray}
when external packets arrive at the source node $\texttt{r}$ at rate $\lambda$.
\end{definition}
\begin{definition} \label{capacity_def}
The broadcast capacity $\lambda^*$ of a network is  the supremum of all arrival rates $\lambda$ for which there exists a broadcast policy $\pi \in \Pi$ of rate $\lambda$.
\end{definition}
In the slotted-time model, the broadcast capacity $\lambda^*$ of a network  $\mathcal{G}$ follows from the Edmonds' tree-packing theorem \cite{edmonds}, and is given by the following:
\begin{eqnarray} \label{capacity_eqn}
\lambda^*=\min_{ \texttt{t} \in V\setminus \{\texttt{r}\}} \text{Max-Flow}(\texttt{r}\to \texttt{t}) \hspace{10pt} \text{per slot,}
\end{eqnarray}
where $\text{Max-Flow}(\texttt{r}\to \texttt{t})$ denotes the maximum value of \emph{flow} that can be feasibly sent from the node $\texttt{r}$ to the node $\texttt{t}$ in the graph $\mathcal{G}(V,E)$ \cite{cormen2009introduction}. Edmonds' theorem also implies that there exist $\lambda^*$ edge-disjoint arborescences \footnote{An arborescence is a directed graph such that there is a unique directed path from the root $\texttt{r}$ to all other vertices in it. Thus, an arborescence is a directed form of a rooted tree. From now onwards, the terms arborescence and directed spanning tree (or simply, spanning tree) will be used interchangeably.} or directed spanning trees, rooted at $\texttt{r}$ in the graph. By examining the flow from the source to every node and using \eqref{capacity_eqn}, it follows that  by sending unit flow over each edge-disjoint tree, we may achieve the capacity $\lambda^*$. \\
As an illustration, consider the graph shown in Figure \ref{diamond_network}. It follows from Eqn. \eqref{capacity_eqn} that the broadcast capacity of the graph is $\lambda^*=2$. Edges belonging to a set of two edge-disjoint spanning trees $\mathcal{T}_1$ and $\mathcal{T}_2$ are shown in blue and red in the figure.  \\
The following lemma establishes the equivalence of the \emph{mini-slot} model and the \emph{slotted-time} model in terms of broadcast-capacity. 
\begin{framed}
\begin{lemma}[\textbf{Invariance of Capacity}]\label{invariance}
The broadcast capacity $\lambda^*$ is the same for both the mini-slot and the slotted-time model and is given by Eqn. \eqref{capacity_eqn}.
\end{lemma}
\end{framed}
\begin{proof}
See Appendix \eqref{invariance_proof}
\end{proof}

\section{A Throughput-Optimal Broadcast Policy \Large{$\pi^*$}} \label{optimal_policy}
In this section we design a throughput-optimal broadcast algorithm $\pi^* \in \Pi$, for networks with arbitrary topology. This algorithm is of \emph{Max-weight} type and is reminiscent of the famous back-pressure policy for the corresponding unicast problem \cite{tassiulas}.  However, because of packet duplications, the usual per-node queues cannot be defined, unlike the unicast case. We get around this difficulty by defining certain virtual-queues, corresponding to subsets of nodes. We show that a scheduling policy in $\Pi^*$, that \emph{stochastically stabilizes} these virtual queues for all arrival rates $\lambda  < \lambda^*$, constitutes a throughput-optimal broadcast policy. Based on this result, we derive a Max-Weight policy $\pi^*$, by minimizing the drift of a quadratic Lyapunov function of the virtual queues.

\subsection{Definitions and Notations} 
To describe our proposed algorithm, we first introduce the notion of  \emph{reachable sets} and \emph{reachable sequence of sets} as follows. 
 \begin{definition}[\textbf{Reachable Set}]
 A subset of vertices $F \subset V$ is said to be \emph{reachable} if the induced graph \footnote{For a graph $\mathcal{G}(V,E)$ and a vertex set $F\subset V$, the induced graph $F(\mathcal{G})$ is defined as the sub-graph containing only the vertices $F$  with the edges whose both ends lie in the set $F$.} $F(\mathcal{G})$ contains a directed arborescence, rooted at source $\texttt{r}$, which spans the node set $F$.
 \end{definition}
 In other words, a subset of vertices $F\subset V$ is reachable if and only if there is a broadcast policy such that, a packet $p$ may be duplicated exactly at the subset $F$ in its course of broadcast. Note that the set of all reachable sets may be strict subset of the set of all subsets of vertices. This is true because all reachable sets, by definition, must contain the source node $\texttt{r}$.
 
In fact, we may completely describe the trajectory of a packet during its course of broadcast, using the notion of \emph{Reachable Sequences}, defined as follows:

 \begin{definition}[\textbf{Reachable Sequence}]
 An ordered sequence of $n-1$ (reachable set, edge) tuples $\{(F_j,e_j), j=1,2,\ldots, n-1\}$ is called a \emph{Reachable Sequence} if the following properties hold:
 \begin{itemize}
 \item $F_1=\{\texttt{r}\}$ and for all $j=1,2,\ldots, n-1$: 
\item $F_j \subset F_{j+1}$
\item  $|F_{j+1}|=|F_j|+1$. 
\item $e_j=(a,b)\in E :a\in F_j, b\in F_{j+1}\setminus F_j$
 \end{itemize}
 $\mathcal{F}$ is defined to be the set of all reachable sequences. 
 \end{definition}

 \begin{figure}
\begin{center}
\includegraphics[scale=1]{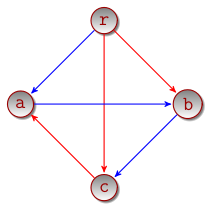}
\end{center}
\caption{\small{\normalfont{The four-node diamond network $\mathcal{D}_4$}}}
\label{diamond_network}
\end{figure}

 A reachable sequence denotes a valid sequence of transmissions for broadcasting a particular packet to all nodes, where the $j$\textsuperscript{th} transmission of a packet takes place across the edge $e_j, j=1,2 \ldots, n-1$. By definition, every reachable set must belong  to at least one reachable sequence. A trivial upper-bound on $|\mathcal{F}|$ is $n^{2n}$.
An example illustrating the notions of reachable sets and reachable sequences for a simple graph is provided below.
  \begin{framed}
 \textbf{Example:}
 Consider the graph shown in Figure \ref{diamond_network}. A reachable sequence for this graph is given by $\mathcal{S}$ below:
 \begin{eqnarray*}
 	\mathcal{S}= \{(\{\texttt{r}\}, \texttt{ra}),(\{\texttt{r,a}\}, \texttt{ab}), (\{\texttt{r,a,b}\}, \texttt{bc})\}
 \end{eqnarray*}
 This reachable sequence is obtained by adding nodes along the tree with blue edges in Figure \ref{diamond_network}. Similarly, an example of a reachable set $F$ in this graph is 
 \begin{eqnarray*}
 	F=\{\texttt{r,a,b}\}
 \end{eqnarray*}
 \end{framed}
 For a reachable set $F$, define its out-edges $\partial^+F$ and in-edges $\partial^-F$ as follows: 
\begin{eqnarray}
\partial^+F=\big\{(a,b)\in E: a \in F, b \notin F \big\}\\
\partial^-F=\big\{(a,b)\in E: a \in F, b \in F \big\}
\end{eqnarray}
For an edge $e=(a,b) \in \partial^+F$, define 
\begin{eqnarray}
	F+e=F \cup \{b\}
\end{eqnarray}
Similarly, for an edge $e=(a,b)\in \partial^-F$, define
\begin{eqnarray}
F\setminus \{e\} = F \setminus \{b\}
\end{eqnarray}
For a sequence of random variables $\{X_n\}_{1}^{\infty}$ and another random variable $X$, defined on the same probability space, by the notation $X_n \stackrel{p}{\implies} X$ we mean that the sequence of random variables $\{X_n\}_{1}^{\infty}$ converges \emph{in probability} to the random variable $X$ \cite{durrett2010probability}. 
 \subsection{System Dynamics}
Consider any broadcast policy $\pi$ in action. For any reachable set $F\subsetneq V$, denote the number of packets, replicated \emph{exactly} at the vertex-set $F$ at mini-slot $t$, by $Q_F(t)$. A packet $p$, which is replicated exactly at the set $F$ by time $t$, is called a \emph{class-$F$ packet}. Hence, at a given time $t$, the reachable sets $F\in \mathcal{F}$ induce a disjoint partition of all the packets in the network. The variable $Q_F(t)$ denotes the number of packets in the partition corresponding to the reachable set $F$. \\
Because of our mini-slot model, a class-$F$ packet can make a transition only to a class $F+e$ (where $e\in \partial^+ F$) during a mini-slot. Let the rate allocated to the edge $e$, for transmitting a class-$F$ packet at time $t$, be denoted by $\mu_{e,F}(t)$\footnote{Note that $\mu_{e,F}(t)$ and consequently, $Q_F(t)$ depends on the algorithm $\pi$ in use and should be denoted by $\mu^\pi_{e,F}(t)$ and $Q^\pi_F(t)$. Here we drop the superscript $\pi$ to simplify notation.}. Here $\mu_{e,F}(t)$ is a binary-valued control variable, which assumes the value $1$ if the edge $e$ (if active) is allocated to transmit a class-$F$ packet at  mini-slot $t$.  The allocated rates are constrained by the underlying random edge-activation process $\{S(t)\}_{0}^{\infty}$. In particular, $\mu_{e,F}(t)$ is zero unless $S(t)=e$. \\
In the following we argue that, for any reachable set $F$, the variable $Q_F(t)$ satisfies following one-step queuing-dynamics (Lindley recursion) \cite{lindley1952theory}:

\begin{eqnarray} \label{dynamics}
Q_F(t+1) &\leq& \bigg(Q_F(t)-\sum_{e \in \partial^+ F} \mu_{e,F}(t)\bigg)^+ + \\
&&\sum_{(e,G):e \in \partial^-F,G=F\setminus\{e\}} \mu_{e,G}(t),\hspace{10pt}\forall F\neq \{\texttt{r}\}\label{q1} \nonumber \\
Q_{\{\texttt{r}\}}(t+1) &\leq &  \bigg(Q_{\{\texttt{r}\}}(t) - \sum_{e \in \partial^+(\{\texttt{r}\})}\mu_{e,\{\texttt{r}\}}(t) \bigg)^+ + A(t) \nonumber 
\end{eqnarray}

The dynamics in Eqn. \eqref{dynamics} may be explained as follows: because of the mini-slot model, only one packet can be transmitted in the entire network at any mini-slot. Hence, for any reachable set $F$, the value of the corresponding state-variable $Q_F(t)$ may go up or down by at most one in a mini-slot. Now, $Q_F(t)$ decreases by one when any of the out-edges $e\in \partial^+F$ is activated at mini-slot $t$ and it carries a class-$F$ packet, provided $Q_F(t)>0$. This explains the first term in Eqn. \eqref{dynamics}.  Similarly, the variable $Q_F(t)$ increases by one when a packet in some set $G=F\setminus \{e\}$ (or an external packet, in case $F=\{\texttt{r}\}$), is transmitted to the set $F$ over the (active) edge $e\in \partial^-F$.  This explains the second term in Eqn. \eqref{dynamics}. In the following, we slightly abuse the notation by setting $\sum_{(e,G):e \in \partial^-F,G=F\setminus\{e\}} \mu_{e,G}(t)\equiv A(t)$, when $F=\{\texttt{r}\}$. Thus the system dynamics is completely specified by the first inequality in \eqref{dynamics}, which constitutes a discrete time \textbf{Lindley recursion} \cite{lindley1952theory}. \\
\subsection{Relationship between Stability and Efficiency}
The following lemma 
shows equivalence between system-stability and throughput-optimality for a Markovian policy. 

\begin{framed}
\begin{lemma}[\textbf{Stability implies Efficiency}]\label{stability-efficiency}
A Markovian policy $\pi$, under which the induced Markov Chain $\{\bm{Q}^\pi(t)\}_{0}^{\infty}$ is Positive Recurrent for all arrival rate $\lambda < \lambda^*$, is a throughput optimal broadcast policy.
\end{lemma}
\end{framed}
\begin{proof}
Under the action of a Markovian Policy $\pi$, the total number of packets $D^\pi(T)$ delivered to all nodes by slot $T$ is given by 
\begin{eqnarray*}
	D^\pi(T) = \sum_{t=1}^{T} A(t) -\sum_{F}Q^\pi_F(T) 
\end{eqnarray*}
Hence, the rate of packet broadcast is given by
\begin{eqnarray}
	\liminf_{T\to \infty}\frac{D^\pi(T)}{T}&= &\liminf_{T\to \infty} \bigg(\frac{1}{T}\sum_{t=1}^{T}A(t)- \sum_{F} \frac{Q^\pi_F(T)}{T}\bigg) \nonumber \\
	&\stackrel{p}{\implies}& \lambda - \sum_F\limsup_{T\to \infty}\frac{Q^\pi_F(T)}{T} \label{conv1}\\
	&\stackrel{p}{\implies}& \lambda \label{conv2}
	\end{eqnarray}
Eqn. \eqref{conv1} follows from the Weak Law of Large Numbers for the arrival process. To justify Eqn. \eqref{conv2}, note that for any $\delta >0$ and any reachable set $F$, we have 
\begin{eqnarray}\label{pr1}
	\lim_{T \to \infty} \mathbb{P}\bigg(\frac{Q^\pi_F(T)}{T}>\delta\bigg) =\lim_{T \to \infty} \mathbb{P}\bigg(Q^\pi_F(T) > T\delta \bigg)=0,  
\end{eqnarray}
where the last equality follows from the definition of positive recurrence. Eqn. \eqref{pr1} implies that $\frac{Q^\pi_F(T)}{T} \stackrel{p}{\implies} 0, \forall F$. This justifies Eqn. \eqref{conv2} and proves the lemma.  
\end{proof}
\subsubsection{Stochastic Stability of the Process \large{$\{\bm{Q}(t)\}_{t\geq 1}$}}
Equipped with Lemma \eqref{stability-efficiency}, we now focus on finding a Markovian policy $\pi^*$, which stabilizes the chain $\bm{Q}^{\pi^*}(t)$\footnote{The time-index $t$ denotes time in mini-slots.}. To accomplish this goal, we use the Lyapunov drift methodology \cite{neely2010stochastic}, and derive a dynamic policy $\pi^*$ which minimizes the one-minislot drift of a certain Lyapunov function. We then show that the proposed policy $\pi^*$ has negative drift outside a bounded region in the state-space. Upon invoking the Foster-Lyapunov criterion \cite{wong2012stochastic}, this proves positive recurrence of the chain $\{\bm{Q}(t)\}_{0}^{\infty}$.\\
To apply the scheme outlined above, we start out by defining the following Quadratic Lyapunov Function $L(\bm{Q}(t))$:
\begin{eqnarray}
L(\bm{Q}(t))= \sum_{F} Q_F^2(t), 
\end{eqnarray}
where the sum extends over all reachable sets. Recall that, the r.v. $S(t)$ denotes the currently active edge at the mini-slot $t$. The one-minislot drift  is defined as: 
\begin{eqnarray} \label{drift}
\Delta_t(\bm{Q}(t),S(t)) \equiv L(\bm{Q}(t+1))- L(\bm{Q}(t))\end{eqnarray}
From the dynamics \eqref{dynamics}, we have 
\begin{eqnarray*}
&&Q_F^2(t+1) \leq Q_F^2(t) + \mu_{\max}^2 \\
&-& 2Q_F(t) \bigg(\sum_{e \in \partial^+ F} \mu_{e,F}(t)- \sum_{(e,G):e \in \partial^-F,G=F\setminus\{e\}} \mu_{e,G}(t) \bigg),
\end{eqnarray*}
where $\mu_{\max}=1$ is the maximum capacity of a link per mini-slot. Thus the one mini-slot drift may be upper-bounded as follows:
\begin{eqnarray*}
&&\Delta_t(\bm{Q}(t), S(t)) \leq 2^{n}\mu_{\max}^2\\
&&-2 \sum_{F\subsetneq V} Q_F(t)  \bigg(\sum_{e \in \partial^+ F} \mu_{e,F}(t)- \sum_{(e,G):e \in \partial^-F,G=F\setminus\{e\}} \mu_{e,G}(t) \bigg).
\end{eqnarray*}
 Interchanging the order of summation, we have 
\begin{eqnarray*}
\Delta_t(\bm{Q}(t), S(t)) &\leq& 2^{n}\mu_{\max}^2\\
&&- \sum_{(e,F):e \in \partial ^+ F}\mu_{e,F}(t) \bigg(Q_F(t)-Q_{F+e}(t)\bigg)
\end{eqnarray*}
Taking expectation of both sides of the above inequality with respect to the edge-activation process $S(t)$ and the arrival process $A(t)$, we obtain the following upper-bound on the conditional Lyapunov drift $\Delta_t(\bm{Q}(t))$, defined as follows: 
\begin{eqnarray} \label{MW_term}
&&\Delta_t(\bm{Q}(t))\equiv \mathbb{E}_{S(t)}\Delta_t(\bm{Q}(t), S(t))  \\
&& \leq  2^{n}\mu_{\max}^2-\nonumber \\
&& \sum_{(e,F):e \in \partial ^+ F} \bigg(Q_F(t)-Q_{F+e}(t)\bigg)\mathbb{E}\big(\mu_{e,F}(t)|\bm{Q}(t),S(t)\big) \nonumber 
\end{eqnarray}
Due to the activity constraint, if $S(t)=e$, we must have $\mu_{l,G}(t)=0, \forall l \neq e$, for all reachable sets $G$. In other words, a packet can only be transmitted along the \emph{active} edge for the mini-slot $t$. Eqn. \eqref{MW_term} immediately leads us to Algorithm \ref{exp-Q}, which is obtained by minimizing the right hand side of the above upper-bound point-wise.
For a reachable set $F$ and an out-edge $e \in \partial^+F$, define the weight 
\begin{eqnarray} \label{wt_def}
w_{F,e}(t) =Q_F(t)-Q_{F+e}(t)
\end{eqnarray}
\begin{algorithm}
At each mini-slot $t$, the network-controller observes the state-vector $\bm{Q}(t)$ and the currently active edge $S(t)=e=(i,j)$ and executes the following steps
\begin{algorithmic}[1]
\STATE Compute all reachable sets $F$ such that $e \in \partial^+ F$.
 \STATE Transmit a class-$F$ packet over the edge $e$, such that the corresponding weight $w_{F,e}(t)=Q_F(t)-Q_{F+e}(t)$ is positive and achieves the maximum over all such reachable sets $F$ under consideration in step 1. 
 \STATE Idle, if no such $F$ exists. 
 \caption{A Dynamic Broadcast Policy $\pi^*$}
 \label{exp-Q}
\end{algorithmic}
\end{algorithm}

We now state the main theorem of this paper.
\begin{framed} 
\begin{theorem}[Throughput-Optimality of $\pi^*$]\label{optimality_thm}
	The dynamic policy $\pi^*$ is a throughput-optimal broadcast policy for any network with arbitrary topology.
\end{theorem}
\end{framed}
\begin{proof}
	See Appendix \eqref{optimality_proof}.
\end{proof}
\section{A Multi-Class Broadcasting \\
Heuristic} \label{heuristic_section}
We note that, the  policy $\pi^*$ makes dynamic routing and scheduling decision for each individual packets, based on the current network-state information $\bm{Q}(t)$. In particular, its operation does not depend on the global topology information of the network. This robustness property makes the policy $\pi^*$ suitable for use in mobile adhoc wireless networks (MANET), where the underlying topology may change frequently. However, a potential difficulty in implementing the policy $\pi^*$ is that, one needs to maintain a state-variable $Q_F(t)$, corresponding to each reachable set $F$, and keep track of the particular reachable set $F_p(t)$ to which packet $p$ belongs. For large networks, without any additional structure in the scheduling policy, maintaining such detailed state-information is quite cumbersome. To alleviate this problem, we next propose a heuristic policy which combines the Max-weight scheduling algorithm designed for $\pi^*$, with the novel idea of \emph{in-class in-order delivery scheme}. The introduction of class-based in-order delivery imposes additional structure in the packet scheduling, which in turn, substantially reduces the complexity of the state-space. \\
\paragraph*{Motivation} To motivate the heuristic policy, we begin with a simple policy-space $\Pi^{\mathrm{in-order}}$, first introduced in \cite{sinha_DAG} for throughput-optimal broadcasting in wireless Directed Acyclic Graphs (DAG). In the space $\Pi^{\mathrm{in-order}}$, the packets are delivered to nodes according to their order of arrival at the source. Unfortunately, as shown in \cite{sinha_DAG}, although $\Pi^{\mathrm{in-order}}$ is sufficient for achieving throughput-optimality in a DAG, it is not necessarily throughput-optimal for arbitrary networks, containing directed cycles. To tackle this problem, we generalize the idea of in-order delivery by proposing a $k$-class policy-space $\Pi_k^{\mathrm{in-order}}, k\geq 1$, which generalizes the space $ \Pi^{\mathrm{in-order}}$. In this space, the packets are separated in $k$ classes. The in-order delivery constraint is imposed in each class but not across classes. Thus, in $\Pi_k^{\mathrm{in-order}}$, the scheduling constraint of $ \Pi^{\mathrm{in-order}}$ is  relaxed by requiring that packets belonging to \emph{each individual class} be delivered to nodes according to their order of arrival at the source. However, the space $\Pi_k^{\mathrm{in-order}}$ does not impose any such restrictions on packet-delivery from different classes. Combining it with the max-weight scheduling scheme, designed earlier for the throughput-optimal policy $\pi^*$, we propose a multi-class heuristic policy $\pi_k^H \in \Pi_k^{\mathrm{in-order}}$ which is \emph{conjectured} to be throughput-optimal for large-enough number of classes $k$. Extensive numerical simulations have been carried out to support this conjecture.  \\
The following section gives detailed description of this heuristic policy, outlined above.
\subsection{The In-order Policy-Space \large{$\Pi^{\mathrm{in-order}}$}}
Now we formally define the policy-space $\Pi^{\mathrm{in-order}}$:
\begin{definition}[Policy-Space $\Pi^{\mathrm{in-order}}$ \cite{sinha_DAG}]
A broadcast policy $\pi$ belongs to the space $\Pi^{\mathrm{in-order}}$ if all incoming packets at the source $\texttt{r}$ are serially indexed $\{1,2,3, \ldots\}$, according to their order of arrivals and a node $i \in V$ is allowed to receive a packet $p$ at time $t$ only if the node $i$ has received the packets $\{1,2,\ldots, p-1\}$ by time $t$. 
\end{definition}
As a result of the \emph{in-order} delivery property of policies in the space $\Pi^{\mathrm{in-order}}$, it follows that the configuration of the packets in the network at time $t$ may be completely represented by the $n$-dimensional vector $\bm{R}(t)$, where $R_i(t)$ denotes the highest index of the packet received by node $i \in V$ by time $t$. We emphasize that this succinct representation of network-state is valid only under the action of the policies in the space $\Pi^{\mathrm{in-order}}$, and is not necessarily true in the general policy-space $\Pi$. \\
Due to the highly-simplified state-space representation, it is natural to try to find efficient broadcast-policies in the space $\Pi^{\mathrm{in-order}}$ for arbitrary network topologies. It is shown in \cite{sinha_DAG} that if the underlying topology of the network is restricted to DAGs, the space $\Pi^{\mathrm{in-order}}$ indeed contains a throughput-optimal broadcast policy. However, it is also shown that the space $\Pi^{\mathrm{in-order}}$ is not rich-enough to achieve broadcast capacity in networks with arbitrary topology. We re-state the following proposition in this connection.
\begin{framed}
\begin{proposition} \label{limitation}
\emph{(}\textsc{Throughput-limitation of the space $\Pi^{\mathrm{in-order}}$ \cite{sinha_DAG}} \emph{)}
There exists a network $\mathcal{G}$ such that, no broadcast-policy in the space $\Pi^{\mathrm{in-order}}$ can achieve the broadcast-capacity of $\mathcal{G}$. 
\end{proposition}
\end{framed}
 The proof of the above proposition is given in \cite{sinha_DAG}, where it is shown that no broadcast policy in the space $\Pi^{\mathrm{in-order}}$ can achieve the broadcast-capacity in the diamond-network $\mathcal{D}_4$, shown in Figure \ref{diamond_network}. 
 \subsection{The Multi-class Policy-Space \large{$ {\Pi}_k^{\mathrm{in-order}}$}}
 To overcome the throughput-limitation of the space $\Pi^{\mathrm{in-order}}$, we propose the following generalized policy-space  $ {\Pi}_k^{\mathrm{in-order}}, k\geq 1$, which retains the efficient representation property of the space $\Pi^{\mathrm{in-order}}$.
 \begin{definition}[Policy-Space $\Pi_k^{\mathrm{in-order}}$ ]
 A broadcast policy $\pi$ belongs to the space $\Pi_k^{\mathrm{in-order}}$ if the following conditions hold:
 \begin{itemize} 
 \item There are $k$ distinct ``classes". 
 \item A packet, upon arrival at the source, is labelled with any one of the $k$ classes, uniformly at random. This label of a packet remains fixed throughout its course of broadcast.
 \item Packets belonging to each individual class $j\in [1,\ldots, k]$, are serially indexed $\{1,2,3, \ldots\}$ according to their order of arrival.
 \item A node $i \in V$ in the network is allowed to receive a packet $p$ from class $j$ at time $t$, only if the node $i$ has received the packets $\{1,2,\ldots, p-1\}$ from the class $j$ by time $t$ . 
 \end{itemize}
 \end{definition}
 In other words, in the policy-space $\Pi_k^{\mathrm{in-order}}$, packets belonging to each individual class $j \in [1,\ldots, k]$ are delivered to nodes in-order. It is also clear from the definition that
 \begin{eqnarray*}
  \Pi_1^{\mathrm{in-order}}=\Pi^{\mathrm{in-order}} 
  \end{eqnarray*}
  Thus, the space $\{\Pi_k^{\mathrm{in-order}}, k\geq 1\}$ generalizes the space $\Pi^{\mathrm{in-order}}$. 
  \paragraph*{State-Space representation under $\Pi_k^{\mathrm{in-order}}$}
  Since each class in the policy-space $\Pi_k^{\mathrm{in-order}}$ obeys the in-order delivery property, it follows that the network-state at time $t$ is completely described by the $k$-tuple of vectors $\{\bm{R}^j(t), 1\leq j\leq k\}$, where $R^j_i(t)$ denotes the highest index of the packet received by node $i \in V$ from class $j$ by time $t$. Thus the state-space complexity grows \emph{linearly} with the number of classes used. \\
 Following our development so far, it is natural to seek a throughput-optimal broadcast policy in the space $\Pi_k^{\mathrm{in-order}}$ with a \emph{small} class-size $k$. In contrast to Proposition \eqref{limitation}, the following proposition gives a positive result in this direction. 
 \begin{framed}
 \begin{proposition} \label{hope}
 \emph{(}\textsc{Throughput-Optimality of the space $\Pi_k^{\mathrm{in-order}}, k\geq n/2$}\emph{)} For every network $\mathcal{G}$, there exists a throughput-optimal broadcast policy in the policy-space $\Pi_k^{\mathrm{in-order}}$ where $k\geq n/2$. 
 \end{proposition} 
 \end{framed}
 The proof of this proposition uses a static policy, which routes the incoming packets along a set of $\lambda^*$ edge-disjoint spanning trees. For a network with broadcast-capacity $\lambda^*$, the existence of these trees are guaranteed by Edmonds' tree packing theorem \cite{edmonds}. Then we show that for any network with unit-capacity edges, its broadcast-capacity $\lambda^*$ is upper-bounded by  $n/2$, which completes the proof. The details of this proof are outlined in Appendix \eqref{hope_proof}. \\
 The policy-class $\Pi_k^{\mathrm{in-order}}$ fixes \emph{intra-class} packet scheduling, by definition. Finally, we need an \emph{inter-class} scheduling policy to resolve contentions among packets from different classes. In the following section, we propose such a scheme.  
\subsection{A Multi-class Heuristic Policy \large{$\pi^H_k \in \Pi^{\mathrm{in-order}}_k$}} 
 
 In this sub-section, we propose a dynamic policy $\pi_k^H \in \Pi_k^{\mathrm{in-order}}$, which uses the same Max-Weight packet scheduling rule, as the throughput-optimal policy $\pi^*$, for \emph{inter-class} packet scheduling. As we will see, the computation of weights and packet scheduling in this case may be efficiently carried out by exploiting the special structure of the space $\Pi_k^{\mathrm{in-order}}$. \\
 We observe that, when the number of classes $k=\infty$ and every incoming packet to the source $\texttt{r}$ joins a new class, the \emph{in-order} restriction of the space $\Pi^{\mathrm{in-order}}_k$ is essentially no longer in effect. In particular, the throughput-optimal policy $\pi^*$ of Section \ref{optimal_policy} belongs to the space $\Pi^{\mathrm{in-order}}_\infty$.  However, we conjecture that the space $\Pi_k^{\mathrm{in-order}}$ is throughput-optimal even when $k=\mathcal{O}(\mathsf{poly}(n))$. Numerical simulation results, supporting this conjecture will be shown subsequently.

 The packet-scheduling algorithm of the policy $\pi^H_k$ may be formally described in the following two parts:
 \paragraph*{Intra-class packet scheduling} As in all policies in the class $\Pi_k^{\mathrm{in-order}}$, when a packet $p$ arrives at the source $\texttt{r}$, it is placed into one of the $k$ classes \emph{uniformly at random}. Packets belonging to any class $c=1,2,\ldots, k$ are delivered to all nodes \emph{in-order} (i.e. the order they arrived at the source $\texttt{r}$). Let the state-variable $R_i^{c}(t)$ denote the number of packets belonging to the class $c$ received by node $i$ up to the mini-slot $t$, $i=1,2,\ldots, n$, $c=1,2,\ldots,k$. As discussed earlier, given the intra-class in-order delivery restriction, the state of the network at the mini-slot $t$ is completely specified by the vector $\big\{\bm{R}^{c}(t),c=1,2,\ldots,k\big \}$.\\
Again, because of the in-order packet-delivery constraint, when an edge $e=(i,j)$ is active at the mini-slot $t$, not all packets that are present at node $i$ and not-present at node $j$ are eligible for transmission. Under the policy $\pi^H_k \in \Pi^{\mathrm{in-order}}_k$, only the next \emph{Head-of-the-Line} (HOL) packet from each class, i.e.,  packet with index $R_j^c(t)+1$ from the class $c$, $c=1,2,\ldots, k$ are eligible to be transmitted to the node $j$, provided that the corresponding packet is also present at node $i$ by mini-slot $t$. Hence, at a given mini-slot $t$, there are at most $k$ contending packets for an active edge. This should be compared with the policy $\pi^*$, in which there are potentially $\mathcal{O}(\exp(n))$ contending packets for an active edge at a mini-slot. \\
\paragraph*{Inter-class packet scheduling}
Given the above intra-class packet-scheduling rule, which follows straight from the definition of the space $\Pi^{\mathrm{in-order}}_k$, we now propose an inter-class packet scheduling, for resolving the contention among multiple contending classes for an active edge $e$ at a mini-slot $t$. For this purpose, we utilize the same Max-Weight scheduling rule, derived for the policy $\pi^*$ (step 2 of Algorithm \ref{exp-Q}).  \\
However, instead of computing the weights $w_{F,e}(t)$ in \eqref{wt_def} for all reachable sets $F$, in this case we only need to compute the weights of the sets $F_c$ corresponding to the HOL packets (if any) belonging to the class $c$. This amounts to a linear number of computations in the class-size $k$. Finally, we schedule the HOL packet from the class $c^*$ having the maximum (positive) weight. By exploiting the structure of the space $\Pi_k^{\mathrm{in-order}}$, the computation of the weights $w_{c}$ can be done in linear-time in the number of classes $k$. It appears from our extensive numerical simulations that $k=\mathcal{O}(m)$ classes suffice for achieving the broadcast capacity in any network.\\
 \paragraph*{Pseudo code} The full pseudo code of the policy $\pi^H_k$ is provided in Algorithm \ref{heuristic_algo}. In lines $4 \ldots 10$, we have used the in-order delivery property of the policy $\pi_k^H$ to compute the sets $F_c$, to which the next HOL packet from the class $c$ belongs. This property is also used in computing the number of packets in the set $G=F_c, F_{c+e}$ in line $14$. Recall that, the variable $Q_G(t)$ counts the number of packets that the reachable set $G$ contains exclusively at mini-slot $t$. These packets can be counted by counting such packets from each individual classes and then summing them up. Again utilizing the in-class in-order delivery property, a little thought reveals that the number of packets $N_G^c(t)$ from class $c$, that belongs exclusively to the set $G$ at time $t$ is given by 
 \begin{eqnarray*}
 N_G^c(t) = \bigg(\min_{i\in G}R^c_i(t) - \max_{i\in V\setminus G}R^c_i(t)\bigg)^+
 \end{eqnarray*}
Hence,
\begin{eqnarray*}
Q_G(t) = \sum_{c=1}^{k}  N_G^c(t),
\end{eqnarray*}
which explains the statement in line $14$. In line $17$, the \emph{weights} corresponding to the HOL packets of each class is computed according to the formula \eqref{wt_def}.  Finally, in line $19$, the HOL packet with the highest positive weight is transmitted across the active edge $e$. The per mini-slot complexity of the policy $\pi^H_k$ is $\mathcal{O}(nk)$.

\begin{algorithm}  
\label{multiclass}
\caption{The Multi-class Scheduling Policy $\pi_k^H$ }\label{heuristic_algo}
At each mini-slot $t$, the network-controller observes the state-variables $\{R_{j}^c(t), j\in V, c=1,2,\ldots, k\}$, the currently active edge $S(t)=e=(i,j)$ and executes the following steps
\begin{algorithmic}[1]
\FORALL {classes $c=1:k$} 
\STATE {\texttt{// Determine the index of the next \textbf{in-order} \\//  (HOL) packet $p_c$ from the class $c$ for node $j$ }}
\STATE $p_c\gets R_j^c(t)+1$.
\STATE { \texttt{// Find the subset $F_c \subset V$ where the packet $p_c$ is currently present:}}
\STATE       $F_c\gets \phi$\\ 
       \FORALL {node $i=1:n$}
          \IF {$R_i^{c}(t)\geq p_c$}
             \STATE $F_c\gets F_c \cup \{i\}$
          \ENDIF
       \ENDFOR 
\STATE $F_{c+e}=F_c \cup \{j\}$   
\STATE {\texttt{// Determine $Q_{F_c}(t)$ and $Q_{F_c+e}(t)$ } }
        \FORALL {$G=F_c, F_{c+e}$}
        \STATE $Q_G(t)\gets \sum_{c=1}^{k}\bigg(\min_{i\in G}R^c_i(t) - \max_{i\in V\setminus G}R^c_i(t)\bigg)^+$
        \ENDFOR
\STATE {\texttt{// Compute the weight $w_c$ for packet $p_c$} }   
    \STATE $w_c \gets \big(Q_{F_c}(t)-Q_{F_c+e}(t)\big)$
\ENDFOR 
\STATE Schedule the packet $p^* \in \arg\max_{c} w_c$, when $\max w_c>0$, else idle. 
\end{algorithmic}
\end{algorithm}
\section{Wireless Interference} \label{int_cont}
A wireless network is modeled by a graph $\mathcal{G}(V,E)$, along with a set of subset (represented by the corresponding binary characteristic vector) of edges $\mathcal{M}$, called the set of \emph{feasible activations} \cite{tassiulas}. The structure of the set $\mathcal{M}$ depends on the underlying interference constraint, e.g., under the \emph{primary interference constraint}, the set $\mathcal{M}$ consists of all \emph{matchings} of the graph $\mathcal{G}$ \cite{west2001introduction}.  Any subset of edges $\bm{s} \in \mathcal{M}$ can be activated simultaneously at a given slot. For broadcasting in wireless networks, we first activate a feasible set of edges from $\mathcal{M}$ and then forward packets on the activated edges. \\
Since the proposed broadcast algorithms in sections \ref{optimal_policy} and \ref{heuristic_section} are \emph{Max-Weight} by nature, they extend straight-forwardly to wireless networks with activation constraints \cite{neely2010stochastic}. In particular, from Eqn. \eqref{wt_def}, at each slot $t$, we first compute the weight of each edge, defined as 
	$w_e(t)= \max_{F:e \in \partial^+ F}w_{e,F}(t)$.
Next, we activate the subset of edges $\bm{s}^*(t)$ from the activation set $\mathcal{M}$, having the maximum weight, i.e.,
\begin{eqnarray*}
	\bm{s}^*(t)=\arg \max_{\bm{s} \in \mathcal{M}} \sum_{e \in E} w_e(t)s_e
\end{eqnarray*}
 Packet forwarding over the activated edges remains the same as before. The above activation procedure carries over to the multi-class heuristic $\pi_k^H$ in wireless networks.

\section{Numerical Simulations} \label{simulation_section}
\subsection{Simulating the Throughput-optimal broadcast policy \large{$\pi^*$}} 
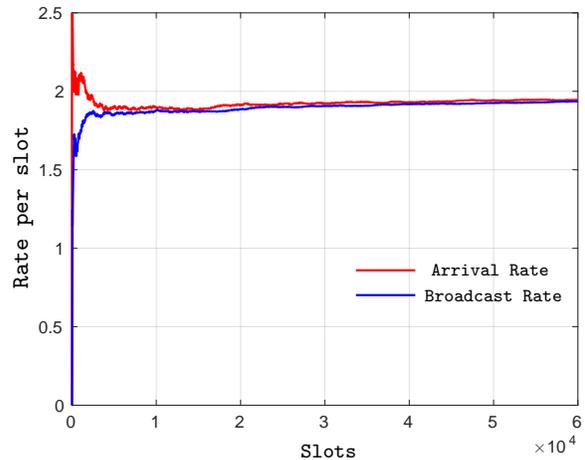
\begin{figure}
\hspace*{-0.5cm}
\begin{overpic}[width=0.49\textwidth] {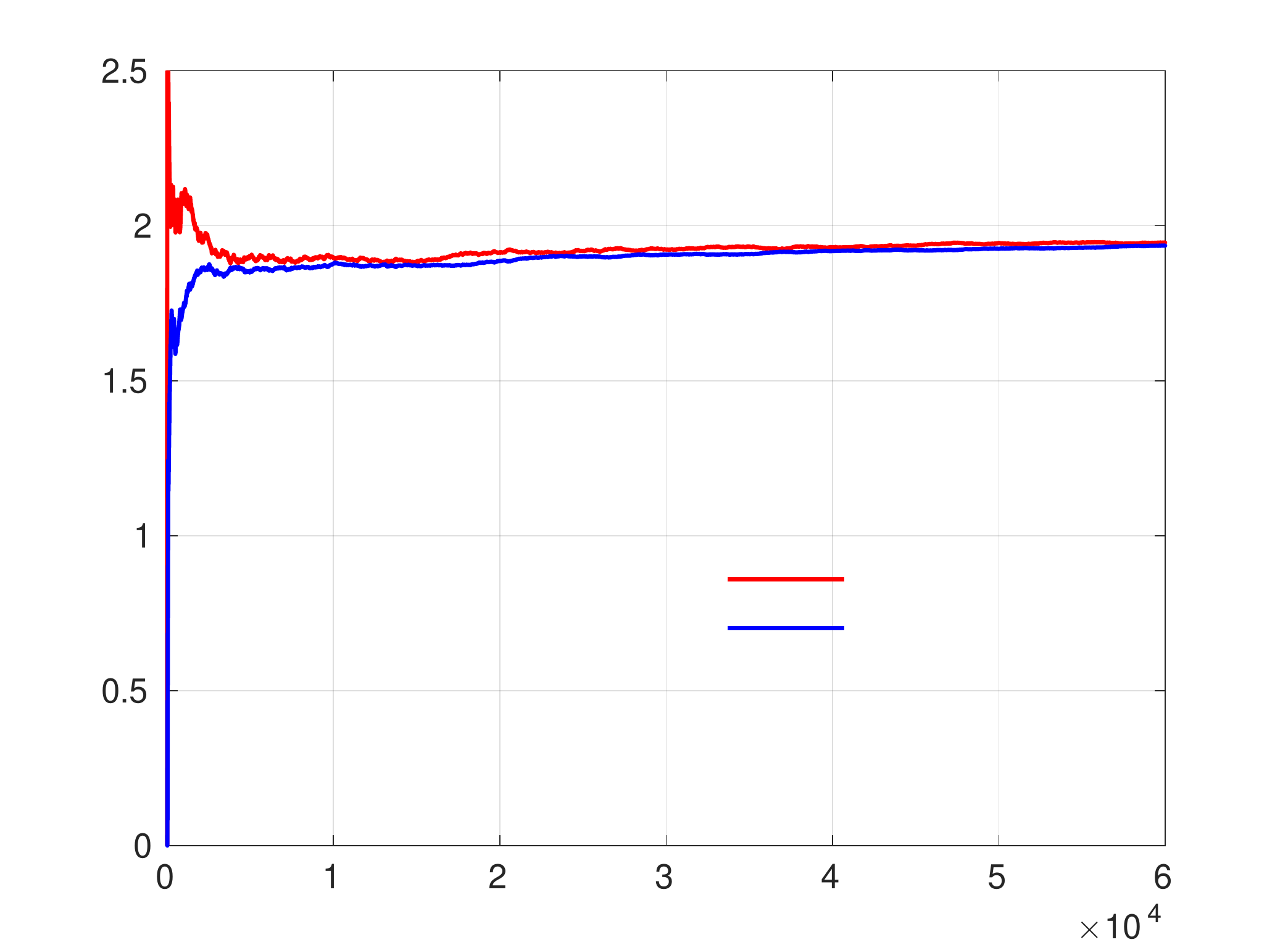}
\put(68,28){\scriptsize{\texttt{Arrival Rate}}}
\put(67,24){\scriptsize{\texttt{Broadcast Rate}}}
\put(48,0){\small{\texttt{Slots}}}
\put(4,26){\rotatebox{90}{\texttt{Rate per slot}}}
\end{overpic}
\caption{Packet Arrival and Broadcast Rate in the Diamond Network in Figure \ref{diamond_network}, under the action of the throughput-optimal policy $\pi^*$ }
\label{broadcast_fig_diamond}
\end{figure}

We simulate the policy $\pi^*$ on the Diamond network $\mathcal{D}_4$, shown in Figure \ref{diamond_network}. The broadcast-capacity of the network is $2$ packets per slot. External packets arrive at the source node $\texttt{r}$ according to a Poisson process of a slightly lower rate of $\lambda=1.95$ packets per slot. A packet is said to be broadcast when it reaches all the nodes in the network. The rate of packet arrival and  packet broadcast by policy $\pi^*$, is shown in Figure \ref{broadcast_fig_diamond}. This plot exemplifies the throughput-optimality of the policy $\pi^*$ for the diamond network. 

\subsection{Simulating the Multi-class Heuristic Policy \large{$\pi_k^{H}$}}
\begin{figure}
\centering
\begin{overpic}[width=0.4\textwidth]{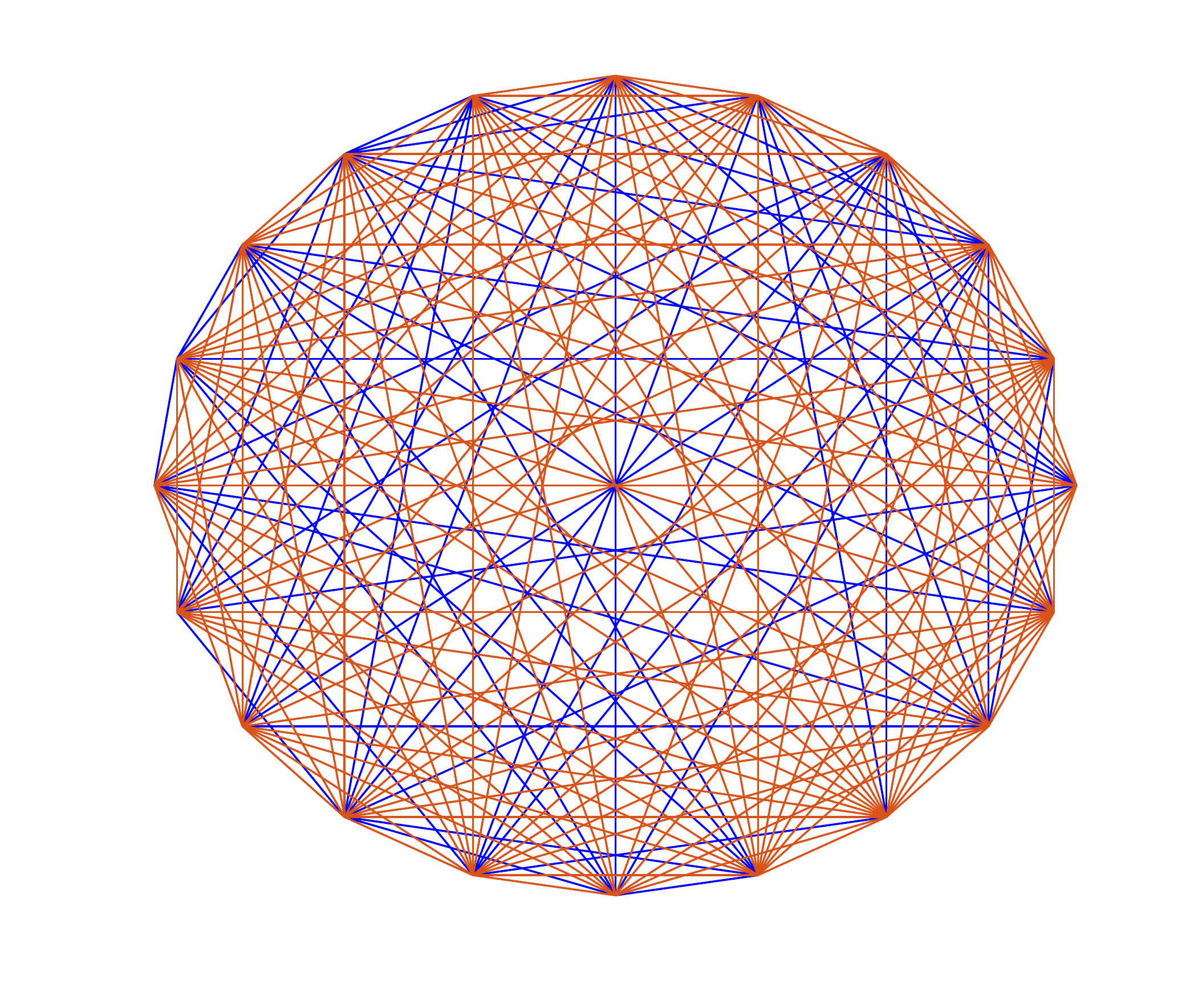}
\put(50,80.5){\small{$6$}}
\put(62.5,78){\small{$5$}}
\put(75.5,72){\small{$4$}}
\put(83.5,64.5){\small{$3$}}
\put(90,54){\small{$2$}}
\put(92,43){\small{$1$}}
\put(90,31.5){\small{$20$}}
\put(83.5,20){\small{$19$}}
\put(75,13){\small{$18$}}
\put(62.5,7.5){\small{$17$}}
\put(50,5){\small{$16$}}
\put(36,7){\small{$15$}}
\put(22.5,13){\small{$14$}}
\put(14,20.5){\small{$13$}}
\put(8,31){\small{$12$}}
\put(6,42){\small{$11$}}
\put(8,54){\small{$10$}}
\put(14,64.5){\small{$9$}}
\put(22,72){\small{$8$}}
\put(35,78){\small{$7$}}
\end{overpic}
\caption{\small {A network $\mathcal{G}$ with $N=20$ nodes. The colors of the edges indicate their directions (e.g., \emph{blue edge} $\implies i\to j : i>j$ and vice versa). The broadcast capacity $\lambda^*$ of the network is computed to be $6$, with node $1$ being the source node.}}
\label{fig:ex_network}
\end{figure}

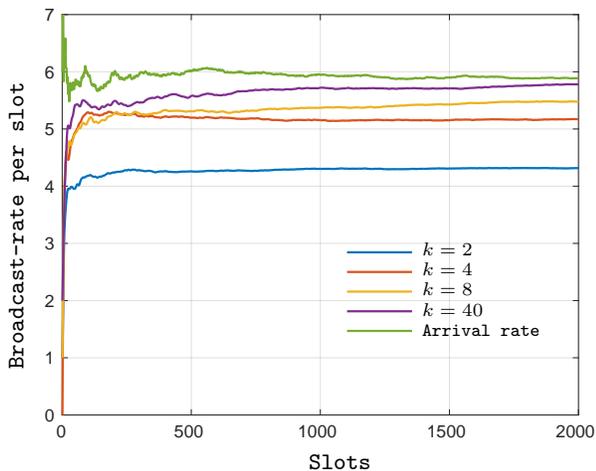
\begin{figure}[!ht]
\centering 
\begin{overpic}[width=0.5\textwidth]{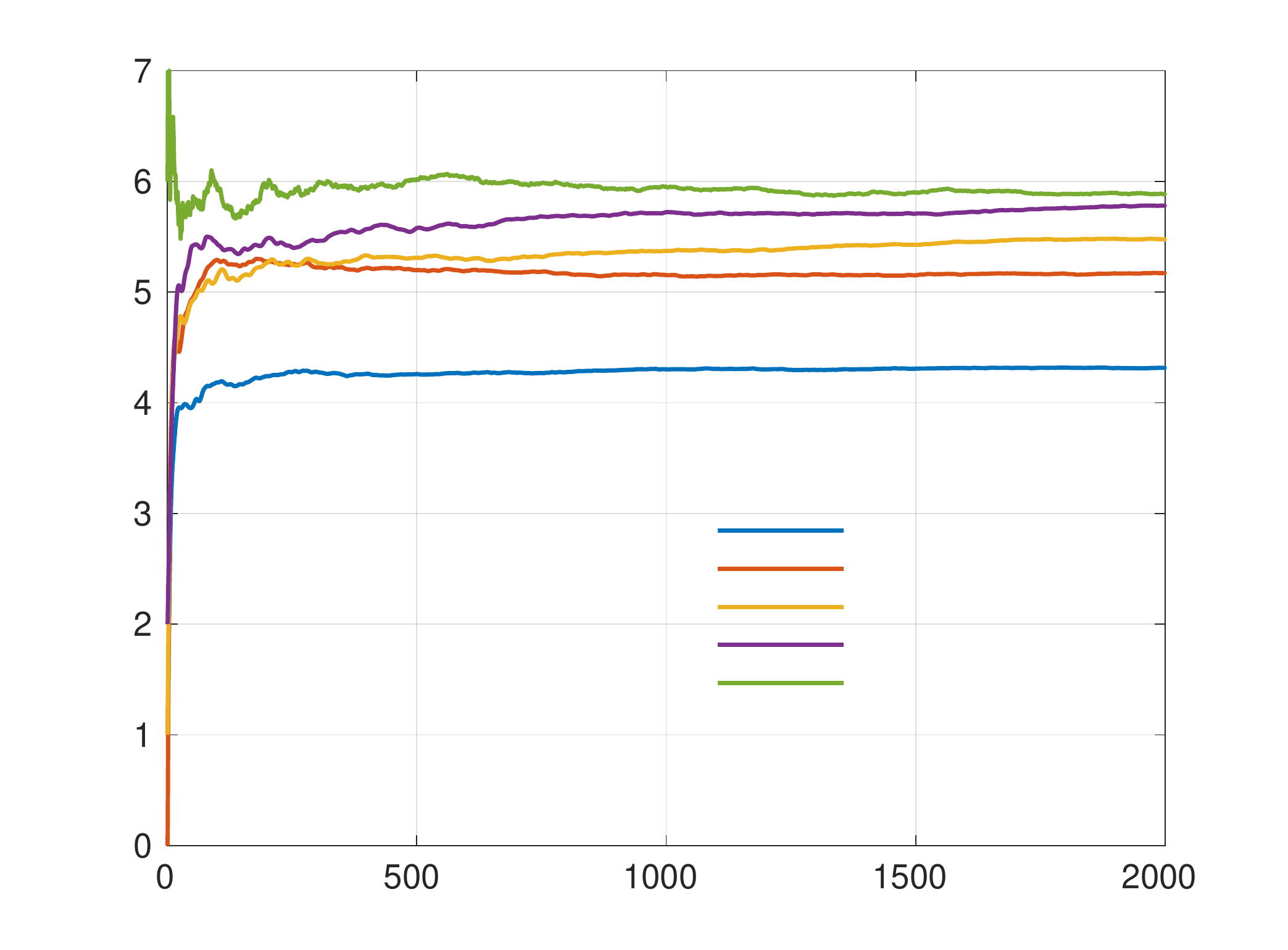}
\put(67,32){\scriptsize{$k=2$}}
\put(67,29){\scriptsize{$k=4$}}
\put(67,26){\scriptsize{$k=8$}}
\put(67,23){\scriptsize{$k=40$}}
\put(67,20){\texttt{\scriptsize{Arrival rate}}}

\put(50,0){\texttt{Slots}}
\put(5,15){\rotatebox{90}{\texttt{Broadcast-rate per slot}}}
\end{overpic}
\caption{Achievable broadcast-rate with the multi-class heuristic broadcast-policies $\pi^H_k$, for $k=2,4,8,40$. The underlying network-topology is given in Figure \ref{fig:ex_network}.}
\label{multi_class_figure}
\end{figure}

The multi-class heuristic policy $\pi^H_k$ has been numerically simulated with $\sim 500$ random networks. We have obtained similar qualitative results in all such instances. One representative sample is discussed here.\\ Consider running the broadcast-policy $\pi^H_k$ on the network shown in Figure \ref{fig:ex_network}, containing $n=20$ nodes and $m=176$ edges. The directions of the edges in this network is chosen arbitrarily.  With node $1$ as the source node, we first compute the broadcast-capacity $\lambda^*$ of this network using Eqn. \eqref{capacity_eqn} and obtain $\lambda^*=6$. External packets arrive at the source node according to a Poisson process, with a slightly smaller rate of $\lambda=5.95$ packets per slot. The rate of broadcast under the multi-class policy $\pi_k^{H}$ for different values of  $k$ is shown in Figure \ref{multi_class_figure}. As evident from the plot, the achievable broadcast rate, obtained by the policy $\pi_k^H$, is non-decreasing in the number of classes $k$. Also, the policy $\pi_k^H$ empirically achieves the broadcast-capacity of the network for a relatively small value of $k= 20$. 

\section{Conclusion and Future Work} \label{conclusion_section}
In this paper we studied the problem of efficient, dynamic packet broadcasting in data networks with arbitrary underlying topology. We derived a throughput-optimal Max-weight broadcast policy that achieves the capacity, albeit at the expense of exponentially many counter-variables. To get around this problem, we next proposed a multi-class heuristic policy which combines the idea of in-order packet delivery with a  Max-weight scheduling, resulting in drastic reduction in the implementation-complexity. The proposed heuristic with polynomially many classes is conjectured to be throughput-optimal. An immediate next step along this line of work would be to prove this conjecture. A problem of practical interest is to find the minimum number of classes $k^*(\epsilon)$ required to achieve a fraction $(1-\epsilon)$ of the capacity. 
\bibliographystyle{abbrv}
\bibliography{MIT_broadcast_bibliography}
\section{Appendix}

\subsection{Proof of Throughput Optimality of \large{$\pi^*$}} \label{optimality_proof}
In this subsection, we show that the induced Markov-Chain $\bm{Q}^{\pi^*}(t)$, generated by the policy $\pi^*$ is positive recurrent, for all arrival rates $\lambda<\lambda^*$ packets per slot. This is proved by showing that the expected one-minislot drift of the Lyapunov function $L(\bm{Q}(t))$ is negative outside a bounded region in the non-negative orthant $\mathbb{Z}_+^{M}$, where $M$ is the dimension of the state-space $\bm{Q}(t)$. To establish the required drift-condition, we first construct an auxiliary stationary randomized policy $\pi^{\text{RAND}}$, which is easier to analyze. Then we bound the one-minislot expected drift of the policy $\pi^*$ by comparing it with the policy $\pi^{\mathrm{RAND}}$. \\
 We emphasize that the construction of the randomized policy $\pi^{\text{RAND}}$ is highly non-trivial, because under the action of the policy $\pi^*$, a packet may travel along an arbitrary tree and as a result, \emph{any} reachable set $F \in \mathcal{F}$ may potentially contain non-zero number of packets. \\ 
For ease of exposition, the proof of throughput-optimality of the policy $\pi^*$ is divided into several parts. \\
\subsubsection{Part I: Consequence of Edmonds' Tree-packing Theorem}
From Edmond's tree-packing theorem \cite{edmonds}, it follows that the graph $\mathcal{G}$ contains $\lambda^*$ edge-disjoint directed spanning trees, \footnote{Note that, since the edges are assumed to be of unit capacity, $\lambda^*$ is an integer. This result follows by combining Eqn. \eqref{capacity_eqn} with the Max-Flow-Min-Cut theorem \cite{cormen2009introduction}. } $\{\mathcal{T}^i\}_{1}^{\lambda^*}$. From Proposition \eqref{invariance} and Lemma \eqref{stability-efficiency}, it follows that, to prove the throughput-optimality of the policy $\pi^*$, it is sufficient to show stochastic-stability of the process $\{\bm{Q}(t)\}_{0}^{\infty}$ for an arrival rate of  $\lambda/m$ per minislot, where $\lambda<\lambda^*$. \\
Fix an arbitrarily small $\epsilon >0$ such that, 
\begin{eqnarray*}
	\lambda \leq \lambda^*-\epsilon
\end{eqnarray*}

Now we construct a stationary randomized policy $\pi^{\mathrm{RAND}}$, which utilizes the edge-disjoint trees $\{\mathcal{T}^i\}_{i=1}^{\lambda^*}$ in a critical fashion.
\subsubsection{Part II: Construction of a Stationary Randomized Policy $\pi^{\text{RAND}}:$}
The stationary randomized policy $\pi^{\mathrm{RAND}}$ allocates rates $\mu_{e,F}(t)$ randomly to different ordered pairs $(e,F)$, for transmitting packets belonging to reachable sets $F$, across an edge $e \in \partial^+F$ \footnote{If $e \notin \partial^+F$, naturally $\mu_{e,F}(t)=0, \forall t$.}. Recall that $\mu_{e,F}(t)$'s are binary variables. Hence, conditioned on the edge-activity process $S(t)=e$, the allocated rates are fully specified by the set of probabilities that a packet from the reachable set $F$ is transmitted across the active edge $e\in \partial^+F$. Equivalently, we may specify the allocated rates in terms of their expectation w.r.t. the edge-activation process (obtained by multiplying the corresponding probabilities by $1/m$). \\
Informally, the policy $\pi^{\mathrm{RAND}}$ allocates most of the rates along the reachable sequences corresponding to the edge-disjont spanning trees $\{\mathcal{T}^i\}_{1}^{\lambda^*}$, obtained in Part I. However, since the dynamic policy $\pi^*$ is not restricted to route packets along the spanning trees $\{\mathcal{T}^i\}_{1}^{\lambda^*}$ only, for technical reasons which will be evident later, $\pi^{\mathrm{RAND}}$  is designed to allocate small amount of rates along other reachable sequences. This is an essential and non-trivial part of the proof methodology. An illustrative example of the rate allocation strategy by the policy $\pi^{\mathrm{RAND}}$ will be described subsequently for the diamond graph $\mathcal{D}_4$ of Figure \ref{diamond_network}. \\
Formally, the rate-allocation by the randomized policy $\pi^{\mathrm{RAND}}$ is given as follows:
\begin{itemize}
\item We index the set of all reachable sequences in a specific order. 

\begin{itemize}
\item The first $\lambda^*$ reachable sequences $\{\zeta^i\}_{i=1}^{\lambda^*}$ are defined as follows: for each edge-disjoint tree $\mathcal{T}^i, i=1,2,\ldots, \lambda^*$ obtained from Part-I, recursively construct a reachable sequence $\zeta^i=\{(F^i_j, e^i_j)\}_{j=1}^{n-1}$, such that the induced sub-graphs
$\mathcal{T}^i(F^i_j)$ are connected for all $j=1,2,\ldots,n-1$.\\ 
In other words, for all $1\leq i \leq \lambda^*$ define $F^i_{1}=\{\texttt{r}\}$ and for all $1\leq j \leq n-2$, the set $F^i_{j+1}$ is recursively constructed from the set $F^i_j$ by adding a node to the set $F^i_j$ while traversing along an edge of the tree $\mathcal{T}^i$. Let the corresponding edge in $\mathcal{T}^i$ connecting the $j+1$ \textsuperscript{th} vertex $F^i_{j+1}\setminus F^i_j$, to the set $F^i_j$, be $e^i_j$. Since the trees $\{\mathcal{T}^i\}_{i=1}^{\lambda^*}$ are edge disjoint, the edges $e^i_j$'s are \textbf{distinct} for all $i=1,2,\ldots, \lambda^*$ and $j=1,2,\ldots, n-1$. The above construction defines the first $\lambda^*$ reachable sequences $\zeta^i=\{F^i_j,e^i_j\}_{j=1}^{n-1}, 1\leq i\leq \lambda^*$.\\
\item In addition to the above, let $\{\zeta^i=(F^{i}_j, e^{i}_j)\}_{j=1}^{n-1}, \lambda^*+1\leq i\leq B $ be the set of all \emph{other} reachable sequence in the graph $\mathcal{G}$, different from the previously constructed $\lambda^*$ reachable sequences. Recall that, $B$ is the cardinality of the set of all reachable  sequences in the graph $\mathcal{G}$. Thus the set of \emph{all} reachable sequences in the graph $\mathcal{G}$ is given by $\bigcup_{i=1}^{B}\zeta^i$. 
\end{itemize}
\item To define the expected allocated rates $\mathbb{E}\mu_{e,F}(t)$, it is useful to first define some auxiliary variables, called \emph{rate-components} $\mathbb{E}\mu^i_{e,F}(t), i=1,2, \ldots, B$, corresponding to each reachable sequence. The rate $\mathbb{E}\mu_{e,F}(t)$ is is simply the sum of the rate-components, as given in Eqn. \eqref{actual_rate}.\\ 
At each slot $t$ and $1 \leq i \leq \lambda^*$, the randomized policy allocates $i$\textsuperscript{th} \emph{rate-component} corresponding to the reachable sequence $\zeta^i=\{e_j^i,F_j^i\}_{j=1}^{n-1}$ according to the following scheme:
\begin{eqnarray} \label{comp_rate}
\mathbb{E}\big(\mu_{e^i_j,F^i_j}^i(t)\big) &=&  1/m- \epsilon (n-j)/n, \nonumber  \\
&& \forall \hspace{2pt} 1\leq j\leq n-1\nonumber\\
&=& 0, \hspace{10pt} \text{o.w.} \label{ex_rate1}
\end{eqnarray}
\item In addition to the rate-allocation \eqref{comp_rate}, the randomized policy $\pi^{\mathrm{RAND}}$ also allocates small amount of rates corresponding to other reachable sequences $\{\zeta^i\}_{\lambda^*+1}^{B}$ according to the following scheme:
  For $\lambda^*+1\leq i\leq B$, the randomized policy allocates $i$\textsuperscript{th} rate-component to the ordered pairs $(e,F)$ as follows: 
\begin{eqnarray}
\mathbb{E}\big(\mu_{e^i_j,F^i_j}^i(t)\big) &=&\frac{\epsilon}{2nB}- \frac{\epsilon}{2nB} \frac{n-j}{n}, \nonumber \\
 && \forall \hspace{2pt}1\leq j \leq n-1, \nonumber \\
&=& 0,\hspace{10pt} \text{o.w.} \label{ex_rate}
\end{eqnarray}
The overall rate allocated to the pair $(e,F)$ is simply the sum of the component-rates, as given below:
\begin{eqnarray}\label{actual_rate}
	\mathbb{E}\mu_{e,F}(t)=\sum_{i=1}^{B} \mathbb{E}\mu^i_{e, F}(t)
\end{eqnarray}
In the following, we show that the above rate-allocation is feasible with respect to the edge capacity constraint.
\end{itemize}
\begin{lemma}[\textbf{Feasibility of Rate Allocation}] \label{feasibility_lemma}
The rate allocation \eqref{actual_rate} by the randomized policy $\pi^{\text{RAND}}$ is feasible.	
\end{lemma}
The reader is referred to Appendix \eqref{feasibility_lemma_proof} for the proof the lemma. An illustrative example for the above randomized rate-allocation scheme is given in Appendix \eqref{example}. 
\subsubsection{Part III: Comparison of drifts under action of policies $\pi^*$ and $\pi^{\mathrm{RAND}}$ } 
Recall that, from  Eqn. \eqref{MW_term} we have the following upper-bound on the one-minislot drift of the Lyapunov function $L(\bm{Q}(t)$, achieved by the policy $\pi^*$:
\begin{eqnarray*}
&&(\Delta^{\pi^*}(\bm{Q}(t)|S(t)) \leq 2^{n}\mu_{\max}^2- \\
&&\sum_{(e,F):e \in \partial ^+ F} \bigg(Q_F(t)-Q_{F+e}(t)\bigg)\mathbb{E}\big(\mu^{\pi^*}_{e,F}(t)|\bm{Q}(t),S(t)\big)
\end{eqnarray*}
Since the policy $\pi^*$, by definition, transmits packets to maximize the weight $w_{F,e}(t)=Q_F(t)-Q_{F+e}(t)$ point wise, the following inequality holds
\begin{eqnarray*}
\sum_{(e,F):e \in \partial ^+ F} \bigg(Q_F(t)-Q_{F+e}(t)\bigg)\mathbb{E}\big(\mu^{\pi^*}_{e,F}(t)|\bm{Q}(t),S(t)\big) \geq \\
\sum_{(e,F):e \in \partial ^+ F} \bigg(Q_F(t)-Q_{F+e}(t)\bigg)\mathbb{E}\big(\mu^{\pi^{\mathrm{RAND}}}_{e,F}(t)|\bm{Q}(t),S(t)\big),
\end{eqnarray*} 
where the randomized rate-allocation $\bm{\mu}^{\pi^{\mathrm{RAND}}}$ is given by Eqn. \eqref{actual_rate}. Noting that $\pi^{\mathrm{RAND}}$ operates independently of the ``queue-states'' $\bm{Q}(t)$ and dropping the super-script $\pi^{\text{RAND}}$ from the control variables $\bm{\mu}(t)$ on the right hand side, we can bound the drift of the policy $\pi^*$ as follows: 

\begin{eqnarray*}
&&(\Delta^{\pi^*}(\bm{Q}(t))|S(t))\\
 &\leq& 2^{n}\mu_{\max}^2- \sum_{(e,F):e \in \partial ^+ F} \bigg(Q_F(t)-Q_{F+e}(t)\bigg)\mathbb{E}\big(\mu_{e,F}(t)|S(t)\big)\\
&=& 2^{n}\mu_{\max}^2- \sum_{F}Q_{F}(t)\bigg(\sum_{e \in \partial^+ F} \mathbb{E}(\mu_{e,F}(t)|S(t))\\
&&- \sum_{(e,G):e \in \partial^-F,G=F\setminus\{e\}} \mathbb{E}(\mu_{e,G}(t)|S(t)) \bigg)\\
\end{eqnarray*}
\begin{eqnarray*}
&\stackrel{(a)}{=}& 2^{n}\mu_{\max}^2- \sum_{F}Q_{F}(t)\bigg(\sum_{e \in \partial^+ F} \big(\sum_{i=1}^{B}\mathbb{E}(\mu^i_{e,F}(t)|S(t))\big)- \\
&&\sum_{(e,G):e \in \partial^-F,G=F\setminus\{e\}} \big(\sum_{i=1}^{B}\mathbb{E}(\mu^i_{e,G}(t)|S(t))\big) \bigg),
\end{eqnarray*}
where in (a) we have used Eqn. \eqref{actual_rate}.\\
 Taking expectation of both sides of the above inequality w.r.t the random edge-activation process $S(t)$ and interchanging the order of summation, we have 
 \begin{eqnarray}\label{expr1}
	\Delta^{\pi^*}(\bm{Q}(t)) &\leq& 2^{n}\mu_{\max}^2- \sum_{F}Q_{F}(t)\sum_{i=1}^{B}\bigg(\sum_{e \in \partial^+ F} \mathbb{E}(\mu^i_{e,F}(t)) \nonumber \\
	&&- \sum_{(e,G):e \in \partial^-F,G=F\setminus\{e\}} \mathbb{E}(\mu^i_{e,G}(t)) \bigg),
\end{eqnarray}
where the rate-components $\bm{\mu}^{i}$ of the randomized policy $\pi^{\mathrm{RAND}}$ are defined in Eqns \eqref{ex_rate1} and \eqref{ex_rate}. \\
Fix a reachable set $F$, appearing in the outer-most summation of the above upper-bound \eqref{expr1}.  Now focus on the $i$\textsuperscript{th} reachable sequence $\zeta^i\equiv \{F_j^i,e_j^i\}_{1}^{n-1}$. We have two cases: \\
\underline{\textbf{Case I: $F \notin \zeta^i$}}\\
Here, according to the allocations in \eqref{ex_rate1} and \eqref{ex_rate}, we have 
\begin{eqnarray*}
	\sum_{e \in \partial^+ F} \mathbb{E}(\mu^i_{e,F}(t))\stackrel{(a)}{=}0, 
	\sum_{(e,G):e \in \partial^-F,G=F\setminus\{e\}} \mathbb{E}(\mu^i_{e,G}(t))\stackrel{(b)}{=}0
\end{eqnarray*}
Where the equality $(a)$ follows from the assumption that $F \notin \zeta^i$ and equality $(b)$ follows from the fact that positive rates are allocated only along the tree corresponding to the reachable sequence $\zeta^i$. Hence, if no rate is allocated to drain packets outside the set $F$, $\pi^{\mathrm{RAND}}$ does not allocate any rate to route packets to the set $F$. 

\underline{\textbf{Case II: $F \in \zeta^i$}}\\
In this case, from Eqns. \eqref{ex_rate1} and \eqref{ex_rate}, it follows that
\begin{eqnarray}
\bigg(\sum_{e \in \partial^+ F} \mathbb{E}(\mu^i_{e,F}(t))- \sum_{(e,G):e \in \partial^-F,G=F\setminus\{e\}} \mathbb{E}(\mu^i_{e,G}(t)) \bigg) \nonumber \\
= \begin{cases}
\frac{\epsilon}{n}, \hspace{30pt}1\leq i\leq \lambda^*\\
	\frac{\epsilon}{2n^2B}, \hspace{5pt}\lambda^*+1\leq i\leq B
	\end{cases}
\end{eqnarray}

By definition, each reachable set is visited by at least one reachable sequence. In other words, there exists at least one $i, 1\leq i \leq B$, such that $F \in \zeta^i$. 
Combining the above two cases, from the upper-bound \eqref{expr1} we conclude that
\begin{eqnarray} \label{drift_expr}
	\Delta^{\pi^*}(\bm{Q}(t))\leq 2^{n}\mu_{\max}^2-\frac{\epsilon}{2n^2B}\sum_F Q_{F}(t),
\end{eqnarray} 
where, the sum extends over \emph{all} reachable sets. The drift is negative, i.e., $\Delta^{\pi^*}(\bm{Q}(t)) < -\epsilon$, when $\bm{Q}_F \in \mathcal{B}^c$, where \begin{eqnarray*}
\mathcal{B}=\bigg\{ (Q_F \geq 0): \sum_{F}Q_F \geq \frac{2n^2B}{\epsilon}(\epsilon+2^n \mu^2_{\max}) \bigg\}
\end{eqnarray*}

Invoking the Foster-Lyapunov criterion \cite{wong2012stochastic}, we conclude that the Markov-Chain $\{\bm{Q}^{\pi^*}(t)\}_{0}^{\infty}$ is positive recurrent. Finally,  throughput-optimality of the policy $\pi^*$ follows from lemma \ref{stability-efficiency}. $\blacksquare$
\newpage 
\subsection{Proof of Lemma \eqref{invariance}}
\begin{proof} \label{invariance_proof}
We prove this lemma in two parts. First, we upper-bound the achievable broadcast rate of the network under any policy in the mini-slot model by the broadcast capacity $\lambda^*(\mathcal{G})$ of the network in the usual slotted model, which is given by Eqn. \eqref{capacity_eqn}. Next, in our main result in section \eqref{optimality_proof}, we constructively show that this rate is achievable, thus proving the lemma. \\ 
Let $\mathcal{C}\subsetneq {V}$ be a non-empty subset of the nodes in the graph $\mathcal{G}$ such that $\texttt{r} \in \mathcal{C}$. Since $\mathcal{C}$ is a strict subset of $V$, there exists a node $i \in V$ such that $i \in \mathcal{C}^c$. Let the set $E(\mathcal{C})$ denote the set of all directed edges $e=(a,b)$ such that $a\in \mathcal{C}$ and $b \notin \mathcal{C}$. Denote $|E(\mathcal{C})|$ by $\text{Cut}(\mathcal{C})$. Using the \textsc{Max-Flow-Min-Cut} theorem \cite{cormen2009introduction}, the broadcast-capacity in the slotted model, given by Eqn. \eqref{capacity_eqn}, may be alternatively represented as
\begin{eqnarray} \label{cut_expr}
\lambda^*= \min_{\mathcal{C}\subsetneq V, \texttt{r} \in \mathcal{C}} \text{Cut}(\mathcal{C})
\end{eqnarray}
 Now let us proceed with the mini-slot model. Since all packets arrived at source $\texttt{r}$ that are received by the node $i$ must cross some edge in the cut $E(\mathcal{C})$, it follows that, under any policy $\pi \in \Pi$, the total number of packets $R_i(t)$ that are received by node $i$ up to mini-slot $t$ is upper-bounded by 
\begin{eqnarray} \label{avg_cut}
R_i(t)\leq \sum_{\tau=1}^{t}  \sum_{e \in E(\mathcal{C})} \mathbbm{1}(S(\tau)=e) = \sum_{e \in E(\mathcal{C})}\sum_{\tau=1}^{t}  \mathbbm{1}(S(\tau)=e)
\end{eqnarray}
 Thus the broadcast-rate $\lambda^\pi_{\text{mini-slot}}$ achievable in the mini-slot model is upper-bounded by  
\begin{eqnarray}\label{ineq1}
&&\lambda^\pi_{\text{mini-slot}} \stackrel{(a)}{\leq} \liminf_{t \to \infty}\frac{R_i(t)}{t}  \stackrel{(b)}{\leq} \liminf_{t \to \infty}\frac{1}{t}\sum_{e \in E(\mathcal{C})}\sum_{\tau=1}^{t} \mathbbm{1}(S(\tau)=e)\nonumber \\
&&=\sum_{e \in E(\mathcal{C})} \lim_{t \to \infty} \frac{1}{t}\sum_{\tau=1}^{t} \mathbbm{1}(S(\tau)=e)\stackrel{(c)}{=} \frac{1}{m}\text{Cut}(\mathcal{C}), \text{  w.p.} 1 
\end{eqnarray} 
Where the inequality (a) follows from the definition of broadcast-rate \eqref{bcdef}, inequality (b) follows from Eqn. \eqref{avg_cut} and finally, the equality (c) follows from the Strong Law of Large Numbers \cite{durrett2010probability}. Since the inequality \eqref{ineq1} holds for any cut $\mathcal{C}\subsetneq C$ containing the source $\texttt{r}$ and any policy $\pi$, from Eqn. \eqref{cut_expr} we have 
\begin{eqnarray}
\lambda^*_{\text{mini-slot}} \leq \lambda^{\pi}_{\text{mini-slot}} \leq \frac{1}{m} \text{Cut}(\mathcal{C}) \leq \frac{1}{m}\lambda^* \text{ per mini-slot} 
\end{eqnarray}
Since according to the hypothesis of the lemma, a slot is identified with $m$ mini-slots, the above result shows that 
\begin{eqnarray}
\lambda^*_{\text{mini-slot}} \leq \lambda^* \text{ per slot} 
\end{eqnarray}
This proves that the capacity in the mini-slot model (per slot) is at most the capacity of the slotted-time model (given by Eqn. \eqref{capacity_eqn}). In section \eqref{optimal_policy}, we show that there exists a broadcast policy $\pi^* \in \Pi$ which achieves a broadcast-rate of $\lambda^*$ packets per-slot in the mini-slot model. This concludes the proof of the lemma.  
\end{proof}

\subsection{Proof of Lemma \eqref{feasibility_lemma}} \label{feasibility_lemma_proof}
\begin{proof} 
The rate allocation \eqref{actual_rate} will be feasible if the sum of the allocated probabilities that an active edge $e$ carries a class-$F$ packet, for all reachable sets $F$, is at most unity. Since  an edge can carry at most one packet per mini-slot, this feasibility condition is equivalent to the requirement that the total expected rate, i.e., $\mathbb{E}\mu_e(t)=\sum_{F}\mathbb{E}\mu_{e,F}(t)$, allocated to an edge $e\in E$ by the randomized policy $\pi^{\mathrm{RAND}}$ does not exceed $\frac{1}{m}$ (the expected capacity of the edge per mini-slot). Since an edge $e$ may appear at most once in any reachable sequence, the total rate allocated to an edge $e$ by the randomized-policy $\pi^{\text{RAND}}$ is upper-bounded by  $\frac{1}{m}- \frac{\epsilon}{n}+ (B-\lambda^*) \frac{\epsilon}{2nB} \leq \frac{1}{m}-\frac{\epsilon}{2n}<1/m$. Hence the rate allocation by the randomized policy $\pi^{\text{RAND}}$ is feasible. 
\end{proof}

\subsection{An Example of Rate Allocation by the Stationary policy $\pi^{\text{RAND}}$}\label{example}
As an explicit example of the above stationary policy, consider the case of the Diamond network $\mathcal{D}_4$, shown in Figure \ref{diamond_network}. The edges of the trees $\{\mathcal{T}^i, i=1,2\}$ are shown in blue and red colors in the figure.  Then the randomized policy allocates the following rate-components to the edges, where the expectation is taken w.r.t. random edge-activations per mini-slot. \\
First we construct a reachable sequence $\zeta^{1}$
consistent with the tree $\mathcal{T}^1$ as follows:
\begin{eqnarray*}
	\zeta^1=\{(\{\texttt{r}\}, \texttt{ra}),(\{\texttt{r,a}\}, \texttt{ab}), (\{\texttt{r,a,b}\}, \texttt{bc})\}
\end{eqnarray*}
Next we allocate the following rate-components as prescribed by $\pi^{\mathrm{RAND}}$:
\begin{eqnarray*}
\mathbb{E}\mu^{1}_{\texttt{ra},\{\texttt{r}\}}(t)&=& 1/6- {3\epsilon}/{4}\\
\mathbb{E}\mu^{1}_{\texttt{ab}, \{\texttt{r,a}\}}(t)&=& 1/6-{2\epsilon}/{4}\\
\mathbb{E}\mu^{1}_{\texttt{bc},\{\texttt{r,a,b}\}}(t)&=&1/6-{\epsilon}/{4}\\
\mathbb{E}\mu^1_{e,F}(t)&=& 0, \hspace{10pt} \text{o.w.} 
\end{eqnarray*}
Similarly for the tree $\mathcal{T}^2$, we first construct a reachable sequence $\zeta^2$ as follows:
\begin{eqnarray*}
	\zeta^2=\{(\{\texttt{r}\}, \texttt{rb}),(\{\texttt{r,b}\}, \texttt{rc}), (\{\texttt{r,b,c}\}, \texttt{ca})\}
\end{eqnarray*}
Then we allocate the following component-rates to the (edge, set) pairs as follows: 
\begin{eqnarray*}
\mathbb{E}\mu^{2}_{\texttt{rb}, \{\texttt{r}\}}(t)&=& 1/6-{3\epsilon}/{4}\\
\mathbb{E}\mu^{2}_{\texttt{rc}, \{\texttt{r,b}\}}(t)&=&1/6-{2\epsilon}/{4}\\
\mathbb{E}\mu^{2}_{\texttt{ca}, \{\texttt{r,b,c}\}}(t) &=&1/6-{\epsilon}/{4}\\
\mathbb{E}\mu^{2}_{e,F}(t)&=&0, \hspace{10pt}
\text{o.w.}
\end{eqnarray*} 
In this example $\lambda^*=2$, thus these two reachable sequence accounts for a major portion of the rates allocated to the edges.
The randomized policy $\pi^{\text{RAND}}$, however, allocates small rates to other reachable sequences too. As an example, consider the following reachable sequence $\zeta^3$, given by 
\begin{eqnarray*}
	\zeta^3= \{(\{\texttt{r}\},\texttt{r}\texttt{a}), (\{\texttt{r,a}\}, \texttt{rb}), (\{\texttt{r,a,b}\}, \texttt{rc}\} 
\end{eqnarray*}
Then, as prescribed above, the randomized policy allocates the following rate-components

\begin{eqnarray*}
	\mathbb{E}\mu^{3}_{\texttt{ra}, \{\texttt{r}\}}(t)&=& \frac{\epsilon}{8B}-\frac{3\epsilon}{32B}\\
\mathbb{E}\mu^{3}_{\texttt{rb}, \{\texttt{r,a}\}}(t)&=&\frac{\epsilon}{8B}-\frac{2\epsilon}{32B}\\
\mathbb{E}\mu^{3}_{\texttt{rc}, \{\texttt{r,a,b}\}}(t) &=&\frac{\epsilon}{8B}-\frac{\epsilon}{32B}\\
\mathbb{E}\mu^{3}_{e,F}(t)&=&0, \hspace{10pt}
\text{o.w.}
\end{eqnarray*}
Here $B$ is the number of all distinct reachable sequences, which is upper-bounded by $4^8$. The rate-components corresponding to other reachable sequences may be computed as above. Finally, the actual expected rate-allocation to the pair $(e,F)$ is given by 
\begin{eqnarray*}
	\mathbb{E}\mu_{e,F}(t) = \sum_{i=1}^{B} \mathbb{E}\mu^i_{e,F}(t) 
\end{eqnarray*}
\subsection{Proof of Proposition \eqref{hope}} \label{hope_proof}
The proof of this proposition is conceptually simplest in the slotted-time model. The argument also applies directly to the mini-slot model. \\
Consider a network $\mathcal{G}$ with broadcast-capacity $\lambda^*$. Assume a slotted-time model. By Edmonds' tree-packing Theorem \cite{edmonds}, we know that there exists $\lambda^*$ number of edge-disjoint directed spanning trees (arborescences) $\{\mathcal{T}_i\}_{1}^{\lambda^*}$ in $\mathcal{G}$, rooted at the source node $\texttt{r}$. Now consider a policy $\pi  \in \Pi_k^{\mathrm{in-order}}$ with $k\geq \lambda^*$ which operates as follows: 
\begin{itemize}
\item An incoming packet is placed in any of the classes $[1,2,$ $3,\ldots, \lambda^*]$, uniformly at random. 
\item Packets in a class $i$ are routed to all nodes in the network \emph{in-order} along the directed tree $\mathcal{T}_i$, where the packets are replicated  in all non-leaf nodes of the tree $\mathcal{T}_i,1\leq i \leq \lambda^*$. 
\end{itemize}
\newpage 
Since the trees are edge-disjoint, the classes do not interact; i.e., routing in each class can be carried out independently. Also by the property of $\mathcal{T}_i$, there is a \emph{unique} directed path  from the source node $\texttt{r}$ to any other node in the network along the edges of the tree $\mathcal{T}_i, 1\leq i \leq \lambda^*$. Thus packets in every class can be delivered to all nodes in the network \emph{in-order} in a pipe-lined fashion with the long-term delivery-rate of $1$ packet per class. Since there are $\lambda^*$ packet-carrying classes, it follows that the policy $\pi\in \Pi_k^{\mathrm{in-order}}$ is throughput-optimal for $k\geq \lambda^*$. \\
Next we show that, $\lambda^*\leq n/2$ for a simple network. Since there exist $\lambda^*$ number of edge-disjoint directed spanning trees in the network, and since each spanning-tree contains $n-1$ edges, we have 
\begin{eqnarray}
\lambda^*(n-1)\leq m 
\end{eqnarray}
Where $m$ is the number of edges in the network. But we have $m \leq n(n-1)/2$ for a simple graph. Thus, from the above equation, we conclude that 
\begin{eqnarray}
\lambda^* \leq n/2.
\end{eqnarray}
This completes the proof of the Proposition. 
%
\end{document}